\documentclass[a4paper,12pt]{article}

\usepackage{ifpdf}

\newif\ifpdf
\ifx\pdfoutput\undefined
  \pdffalse
\else
  \pdfoutput=1
  \pdftrue
\fi

\RequirePackage{xspace} %
\RequirePackage{subfigure} %
\RequirePackage[centertags]{amsmath} %
\RequirePackage{amssymb}
\RequirePackage{wrapfig} %
\RequirePackage{calc} %
\RequirePackage{ifthen}
\RequirePackage{tabularx} %
\RequirePackage{flafter} %
\RequirePackage{fancyhdr} %

\ifpdf
  \RequirePackage[pdftex]{color}%
  \RequirePackage{colortbl}%
  \RequirePackage{array}%
  \RequirePackage[pdftex]{graphicx}

  \RequirePackage[ pdftex, plainpages = false, pdfpagelabels,
                 pdfpagelayout = useoutlines,
                 bookmarks,
                 breaklinks = true,
                 linktocpage,
                 pagebackref,
                 colorlinks = true,
                 linkcolor = blue,
                 urlcolor  = blue,
                 citecolor = blue,
                 anchorcolor = blue,
                 hyperindex = true,
                 hyperfigures
                 ]{hyperref}

\else
  \RequirePackage{color}
  \RequirePackage{colortbl}
   \RequirePackage{array}
  \RequirePackage[dvips]{graphicx}
  \RequirePackage{hyperref}
  \usepackage{rotating}
\fi

\usepackage{makeidx}
\usepackage{setspace}
\usepackage{rotating}
\usepackage{ecltree}
\usepackage{epic}
\usepackage{supertabular}
\usepackage{color}
\usepackage{exscale}
\usepackage{fontenc}
\usepackage{ifthen}
\usepackage{latexsym}
\usepackage{makeidx}
\usepackage{syntonly}
\usepackage{inputenc}
\usepackage{graphicx}
\usepackage{setspace}
\usepackage{caption2}
\usepackage[english]{babel}
\usepackage[square, comma,numbers,sort&compress]{natbib}
\usepackage{boxedminipage}
\usepackage{framed}
\usepackage{longtable}
\usepackage{lscape}
\usepackage{amsmath}
\usepackage[all]{hypcap}

\renewcommand{\sectionmark}[1]%
      {\markright{\thesection\ #1}}

\newcommand{\etal}    {{\it et al}}

\newcommand{\D}   [2]  {\frac{d #1}{d #2}}
\newcommand{\DELE}[2] {\Delta \E_{#1#2}}
\newcommand{\DELK}[2] {\Delta \K_{#1#2}}

\newcommand{\EP}     {\delta}
\newcommand{\EPB}    {\overline{\EP}}
\newcommand{\EPS}    {\EP}
\newcommand{\EPGK}   {\delta^{'}}
\newcommand{\RW}     {\Delta_{r}}
\newcommand{\LTS}    {\tau}
\newcommand{\FWHMe}  {\Delta_{E}}
\newcommand{\EV}     {\Lambda}

\newcommand{\BGCK}   {K_{_{o}}}
\newcommand{\E}      {E}

\newcommand{\Ei}     {\E_{i}}

\newcommand{\Eo}     {\E_{0}}
\newcommand{\Er}     {\E_{r}}
\newcommand{\PKF}    {\rm{g}}
\newcommand{\h}      {h}
\newcommand{\Dirac}  {\hbar}
\newcommand{\SVKi}   {K_{i}}

\newcommand{\iu}    {\textrm{i}}
\newcommand{\B}     {B}
\newcommand{\Bm}    {{\bf{\B}}\index{B-matrix}}
\newcommand{\I}     {I}
\newcommand{\IM}    {{\bf{\I}}}
\newcommand{\K}     {K}
\newcommand{\Km}    {{\bf{\K}}\index{K-matrix}}
\newcommand{\M}     {M}
\newcommand{\Mm}    {{\bf{\M}}\index{M-matrix}}
\newcommand{\Mmax}  {\M_{max}}
\newcommand{\Q}     {Q}
\newcommand{\Qm}    {{\bf{\Q}}\index{Q-matrix}}
\newcommand{\QB}    {{\bf{\Q\B}}\index{QB method}}
\newcommand{\R}     {R}
\newcommand{\Rm}    {{\bf{\R}}\index{R-matrix}}
\newcommand{\Ss}    {S}
\newcommand{\Sm}    {{\bf{\Ss}}\index{S-matrix}}
\newcommand{\STGQB}      {STGQB\index{STGQB}}
\newcommand{\Planck}     {Planck\index{Planck}}

\newcommand{\timdel}     {time-delay\index{Time-Delay}}
\newcommand{\TimDel}     {Time-Delay\index{Time-Delay}}
\newcommand{\TIMEDEL}    {TIMEDEL\index{TIMEDEL}}
\newcommand{\II}         {~{\sc ii}}
\newcommand{\SLP} [3]{$^{#1}$#2$^{\rm{#3}}$}
\newcommand{\SLPJ}[4]{$^{#1}$#2$^{\rm{#3}}_{_{#4}}$}

\doublespace 

\title
 {
\vspace*{3.0cm} \Large{\bf Methods for Analyzing Resonances in Atomic Scattering} \vspace*{3.0cm} \\
\Large{\bf Taha Sochi\footnote{Corresponding author. Email: t.sochi@ucl.ac.uk.}\, and P.J. Storey} \vspace*{1.0cm} \\
\footnotesize{University College London, Department of Physics and Astronomy, Gower Street, London, WC1E 6BT}\vspace*{3.0cm} \\
}

\begin{document}

\maketitle %
\pagenumbering{arabic}

\newpage
\phantomsection \addcontentsline{toc}{section}{Abstract} \noindent
{\noindent \LARGE \bf Abstract} \vspace{0.5cm}\\

Resonances, which are also described as autoionizing or quasi-bound states, play an important role
in the scattering of atoms and ions with electrons. The current article is an overview of the main
methods, including a recently-proposed one, that are used to find and analyze resonances.

\vspace{0.5cm}

Keywords: atomic scattering; resonance; autoionizing states; quasi-bound states; \Rm-matrix;
\Km-matrix method; \QB\ method; time-delay method.

\pagestyle{headings} %
\addtolength{\headheight}{+1.6pt}
\lhead[{Chapter \thechapter \thepage}]%
      {{\bfseries\rightmark}}
\rhead[{\bfseries\leftmark}]%
     {{\bfseries\thepage}} 
\headsep = 1.0cm               

\newpage
\section{Introduction} \label{Introduction}

In atomic scattering, resonance occurs when a colliding continuum electron is captured by an ion to
form a doubly excited state. The resonant state is normally short lived and hence it is permanently
stabilized {\it either} by a radiative decay of the captured electron or an electron in the parent
to a true bound state of the excited core ion {\it or} by autoionization to the continuum with the
ejection of an electron. In most cases the stabilization occurs through the second route, i.e. by
autoionization rather than radiative decay. These excited autoionizing systems leave a distinctive
signature in the cross sections for electron scattering processes. The Auger effect is one of the
early examples of the resonance effects that have been discovered and extensively investigated
\cite{Burke1965, BartschatB1986, SochiSTemp2013}. The reader is advised to consult other references
in the literature of atomic physics (e.g. \cite{Burke1965, Drake2006}) for a general historical
background.

Symmetric and asymmetric line shapes have been proposed to model resonance profiles for different
situations; these profiles include Lorentz, Shore and Fano, as given in Table \ref{ProfilesTable}.
Resonance characteristics; such as their position on the energy axis, area under profile and full
width at half maximum; are usually obtained by fitting the profile of the autoionizing state to a
fitting model such as Lorentz. These characteristics have physical significance; for example the
width of a resonance quantifies the strength of the interaction with the continuum and hence
autoionization probability and lifetime, while the contribution of a resonance to a photoionization
cross section, quantified by the area beneath it, is related to the probability of radiative decay
\cite{Storey1994, Drake2006, SochiEmis2010, SochiSCIIList2013, SochiThesis2012}.

\begin{table} [!h]
\caption{The commonly-used resonance line profiles where $\sigma$ is the photoionization cross
section, $A$ is a proportionality factor with the dimension of area, $\Delta_{E}$ is the full width
at half maximum, $E$ is the energy, $E_{r}$ is the resonance position on the energy line,
$p=\frac{2\left(E-E_{r}\right)}{\Delta_{E}}$, $\alpha$ and $\beta$ are parameters related to the
dipole and Coulomb matrix elements, and $k$ is the Fano asymmetry factor. \label{ProfilesTable}}
\begin{center} \vspace{0.3cm}
\begin{tabular}{|l|l|}
\hline
 Profile \hspace{0.5cm} & Equation\tabularnewline
 \hline
 Lorentz & $\sigma=A\frac{\Delta_{E}^{2}/4}{\left(E-E_{r}\right)^{2}+\Delta_{E}^{2}/4}$\hspace{0.5cm}\tabularnewline
 \hline
 Shore & $\sigma=A\frac{\alpha p+\beta}{p^{2}+1}$\tabularnewline
 \hline
 Fano & $\sigma=A\frac{\left(k+p\right)^{2}}{p^{2}+1}$\tabularnewline
 \hline
\end{tabular}
\par\end{center}
\end{table}

\section{Methods for Investigating Resonances} \label{Methods}

There are several methods for finding and analyzing resonances. In the following sections we
outline three of these methods which are all based on the use of the reactance \Km-matrix of the
\Rm-matrix theory for atomic and molecular scattering calculations. The advantage of this common
approach, which employs the close coupling approximation, is that resonance effects are naturally
delineated, since the interaction between bound and free states is incorporated in the scattering
treatment.

\subsection{QB Method} \label{QBmethod}

A common approach for finding and analyzing resonances is to apply a fitting procedure to the
reactance matrix, \Km, or its eigenphase as a function of energy in the neighborhood of an
autoionizing state. However, fitting the \Km-matrix itself is complicated because the reactance
matrix has a pole at the energy position of the autoionizing state. An easier alternative is to fit
the arc-tangent of the reactance matrix. The latter approach was employed by Bartschat and Burke
\cite{BartschatB1986} in their fitting code RESFIT.

The eigenphase sum is defined by
\begin{equation}\label{EigSum}
    \EPS = \sum_{i=1}^{N} \arctan \EV_{i}
\end{equation}
where $\EV_{i}$ is an eigenvalue of the \Km-matrix and the sum runs over all open channels
interacting with the autoionizing state. The eigenphase sum is normally fitted to a Breit-Wigner
form
\begin{equation}\label{BreWig}
    \EPS = \EPB + \arctan \left(   \frac{\RW}{2\left(\Er - \E\right)}   \right)
\end{equation}
where $\EPB$ is the sum of the background eigenphase and $\RW$ is the resonance width. This
approach was used by Tennyson and Noble \cite{TennysonN1984} in their fitting code RESON
\cite{StibbeT1996}.

In theory, an autoionizing state exhibits itself as a sharp increase by $\pi$ radians in the
eigenphase sum as a function of energy superimposed on a slowly-varying background. However, due to
the finite width of resonances and the background variation over their profile, the increase may
not be by $\pi$ precisely in the actual calculations. A more practical approach then is to identify
the position of the resonance from the energy location where the increase in the eigenphase sum is
at its highest rate by having a maximum gradient with respect to the scattering energy, i.e.
$\left(d\EPS/d\E\right)_{max}$ \cite{TennysonN1984, QuigleyB1996, BusbyBBNS1998}.

The \QB\ method of Quigley and Berrington \cite{QuigleyB1996} is a computational technique for
finding and analyzing autoionizing states that arise in atomic and molecular scattering processes
using eigenphase fitting. The essence of this method is to apply a fitting procedure to the
reactance matrix eigenphase near the resonance position using the analytic properties of the
\Rm-matrix theory. The merit of the \QB\ method over other eigenphase fitting procedures is that it
utilizes the analytical properties of the \Rm-matrix method to determine the variation of the
reactance matrix with respect to the scattering energy analytically. This analytical approach can
avoid possible weaknesses, linked to the calculations of \Km-matrix poles and arc-tangents, when
numerical procedures are employed instead. The derivative of the reactance matrix with respect to
the scattering energy in the neighborhood of a resonance can then be used in the fitting procedure
to identify the energy position and width of the resonance.

The \QB\ method begins by defining two matrices, \Qm\ and \Bm, in terms of asymptotic solutions,
the \Rm-matrix and energy derivatives, such that
\begin{equation}\label{QB}
    \D \Km \E = \Bm^{-1} \Qm
\end{equation}
The gradients of the eigenphases of the \Km-matrix with respect to energy can then be calculated.
This is followed by identifying the resonance position, $\Er$, from the point of maximum gradient
at the energy mesh, and the resonance width, $\RW$, which is linked to the eigenphase gradient at
the resonance position, $\EPGK(\Er)$, by the relation
\begin{equation}\label{QBwidth}
    \RW = \frac{2}{\EPGK(\Er)}
\end{equation}
This equation may be used to calculate the widths of a number of resonances in a first
approximation. A background correction due to overlapping profiles can then be introduced on these
widths individually to obtain a better estimate.

The \QB\ method was implemented in the \STGQB\ code of Quigley and coworkers \cite{QuigleyBP1998}
as an extension to the \Rm-matrix code. It should be remarked that Busby \etal\
\cite{BusbyBBNS1998} have used a similar method for finding and analyzing resonances graphically by
their VisRes program.

\subsection{\TimDel\ Method} \label{TDmethod}

The \TimDel\ method of Stibbe and Tennyson \cite{StibbeT1998} is based on the \timdel\ theory of
Smith \cite{Smith1960} where use is made of the lifetime eigenvalues to locate the resonance
position and identify its width. According to this theory, the \timdel\ matrix \Mm\ is defined in
terms of the scattering matrix \Sm\ by
\begin{equation}\label{TDmatrixS2}
    \Mm = -\iu \, \Dirac \, \Sm^{*} \frac{d \Sm}{d \E}
\end{equation}
where \iu\ is the imaginary unit, $\Dirac$ ($=\h/2\pi$) is the reduced \Planck's constant, and
$\Sm^{*}$ is the complex conjugate of $\Sm$. It has been demonstrated by Smith \cite{Smith1960}
that the eigenvalues of \Mm\ represent the collision lifetimes and the largest of these eigenvalues
corresponds to the longest \timdel\ of the scattered particle. For a resonance, the \timdel\ has a
Lorentzian profile with a maximum precisely at the resonance position. By computing the
energy-dependent \timdel\ from the reactance matrix, and fitting it to a Lorentzian peak shape, the
resonance position can be located and its width is identified.

This method, as implemented in the \TIMEDEL\ program of Stibbe and Tennyson \cite{StibbeT1998},
uses the reactance \Km-matrix as an input, either from a readily-available archived scattering
calculations or from dynamically-performed computations on an adjustable mesh. The \Sm-matrix is
then formed using the relation
\begin{equation}\label{SmatrixEq2}
    \Sm = \frac{\IM + \iu \Km}{\IM - \iu \Km}
\end{equation}
where $\IM$ is the identity matrix. The \timdel\ \Mm-matrix is then calculated from
Equation~\ref{TDmatrixS2}, with numerical evaluation of the \Sm-matrix derivative, and diagonalized
to find the eigenvalues and hence obtain the longest \timdel\ of the incident particle. Approximate
resonance positions are then identified from the energy locations of the maxima in the \timdel\
profile, and the widths are estimated from the Lorentzian fit. On testing the degree of overlapping
of neighboring resonances, \TIMEDEL\ decides if the resonances should be fitted jointly or
separately.

\subsection{K-Matrix Method} \label{Kmethod}

Sochi \cite{SochiThesis2012} and Sochi \& Storey \cite{SochiSCIIList2013} studied the properties of
autoionizing states of the C$^{2+}$+e$^-$ system at energies close to the ionization limit. In
this, and other Be-like systems, there is only one open channel per angular momentum symmetry and
the \Km-matrix is a real scalar. This simplification makes it possible to directly analyze the
poles in \Km\ to derive resonance properties as outlined below.

According to the collision theory of Smith \cite{Smith1960}, \Mm-matrix is related to \Sm-matrix
by Equation~\ref{TDmatrixS2}. Now, a single-channel \Km-matrix with a pole at energy $\Eo$
superimposed on a background $\BGCK$ can be approximated by
\begin{equation}\label{Kmatrix}
    \SVKi = \BGCK + \frac{\PKF}{\Ei - \Eo}
\end{equation}
where $\SVKi$ is the value of the \Km-matrix at energy $\Ei$ and $\PKF$ is a physical parameter
with dimension of energy. In Appendix~A it is demonstrated that in the case of single-channel
scattering the \Mm-matrix is real with a value given by
\begin{equation}\label{mmatrix}
    \M = \frac{-2 \PKF}{(1+ \BGCK^{2})(\E-\Eo)^{2}+2 \BGCK \PKF (\E - \Eo) + \PKF^{2}}
\end{equation}
Using the fact demonstrated by Smith \cite{Smith1960} that the lifetime of the state is the
expectation value of $\M$, it can be shown from Equation~\ref{mmatrix} that the position of the
resonance peak $\Er$ is given by
\begin{equation}\label{Er}
    \Er = \Eo - \frac{\BGCK \PKF}{1 + \BGCK^{2}}
\end{equation}
while the full width at half maximum $\FWHMe$ is given by
\begin{equation}\label{FWHM}
    \FWHMe = \frac{|2 \PKF|}{1 + \BGCK^{2}}
\end{equation}
Complete derivation of the \Km-matrix method is given in Appendix~A.

The two parameters of primary interest are the resonance energy position $\Er$, and the resonance
width $\RW$ which equals the full width at half maximum $\FWHMe$. However, for an energy point
$\Ei$ with a \Km-matrix value $\SVKi$, Equation~\ref{Kmatrix} has three unknowns, $\BGCK$, $\PKF$
and $\Eo$, which are needed to find $\Er$ and $\RW$. Hence, three energy points in the immediate
neighborhood of $\Eo$ are required to identify these unknowns. As the \Km-matrix changes sign at
the pole, the neighborhood of $\Eo$ is located by testing the \Km-matrix value at each point of the
energy mesh for a sign change or a discontinuity evidenced by a sharp change in the gradient.

In practical terms, the method of locating the \Km-matrix poles is as follows. The asymptotic
routine STGF \cite{BerringtonESSS1987, BerringtonEN1995, BadnellRmax2013} in the \Rm-matrix package
was modified by the authors to test for a sign change in the value of \Km\ as it is calculated,
initially using a coarse energy mesh. If a sign change is detected, the new STGF routine goes back
in the energy mesh and defines a new fine mesh over very limited energy range that includes the
sign-change position and outputs the energy points of the fine mesh and the corresponding
\Km-matrix values to be used for finding the resonance parameters.

The modified STGF also writes the \Km-matrix values to a file. In case a sign change is not
detected, a separate code reads the \Km-matrix and tests the slope for a sudden change where poles
could exist. If such a change is detected a new finer energy mesh around the suspected pole is
prepared for a new STGF run. The process is repeated until a sign change is seen in the \Km-matrix
at which point the resonance parameters are computed.

\begin{table} [!h]
\caption{The two C\II\ resonances used for demonstrating the \Km-matrix poles in Figures
\ref{Graph32} and \ref{Graph41}. The columns from left to right are: configuration, level,
experimental energy in wavenumbers (cm$^{-1}$) relative to the ground state, experimental energy in
Rydberg relative to the C$^{2+}$ \SLPJ1Se0\ limit, theoretical energy in Rydberg from \Km-matrix
calculations relative to the C$^{2+}$ \SLPJ1Se0\ limit, full width at half maximum from \Km-matrix
in Rydberg, theoretical energy in Rydberg from \QB\ calculations relative to the C$^{2+}$
\SLPJ1Se0\ limit, and full width at half maximum from \QB\ in Rydberg. The experimental data are
obtained from the National Institute of Standards and Technology (NIST). \label{ResTable}}
 \vspace{-0.2cm}
\begin{center}
{\footnotesize
\begin{tabular}{@{\extracolsep\fill}llllllll@{}}
\hline {Config.} & {Lev.} & {NEEW} & {NEER} & {TERK} & {FWHMK} & {TERQ} & {FWHMQ} \\
\hline
1s$^{2}$2s2p(\SLP3Po)4d     &    \SLPJ4Fo{7/2}  &   219590.76   &   0.208918    &  0.209174    &   5.96E-09    &   0.209174    &   5.96E-09    \\
1s$^{2}$2s2p(\SLP3Po)4d     &    \SLPJ4Po{3/2}  &   220832.15   &   0.220230    &  0.220680    &   5.32E-10    &   0.220680    &   5.32E-10    \\
\hline
\end{tabular}
}
\end{center}
\end{table}

\begin{figure}[!h]
\centering{}
\includegraphics
[scale=0.65] {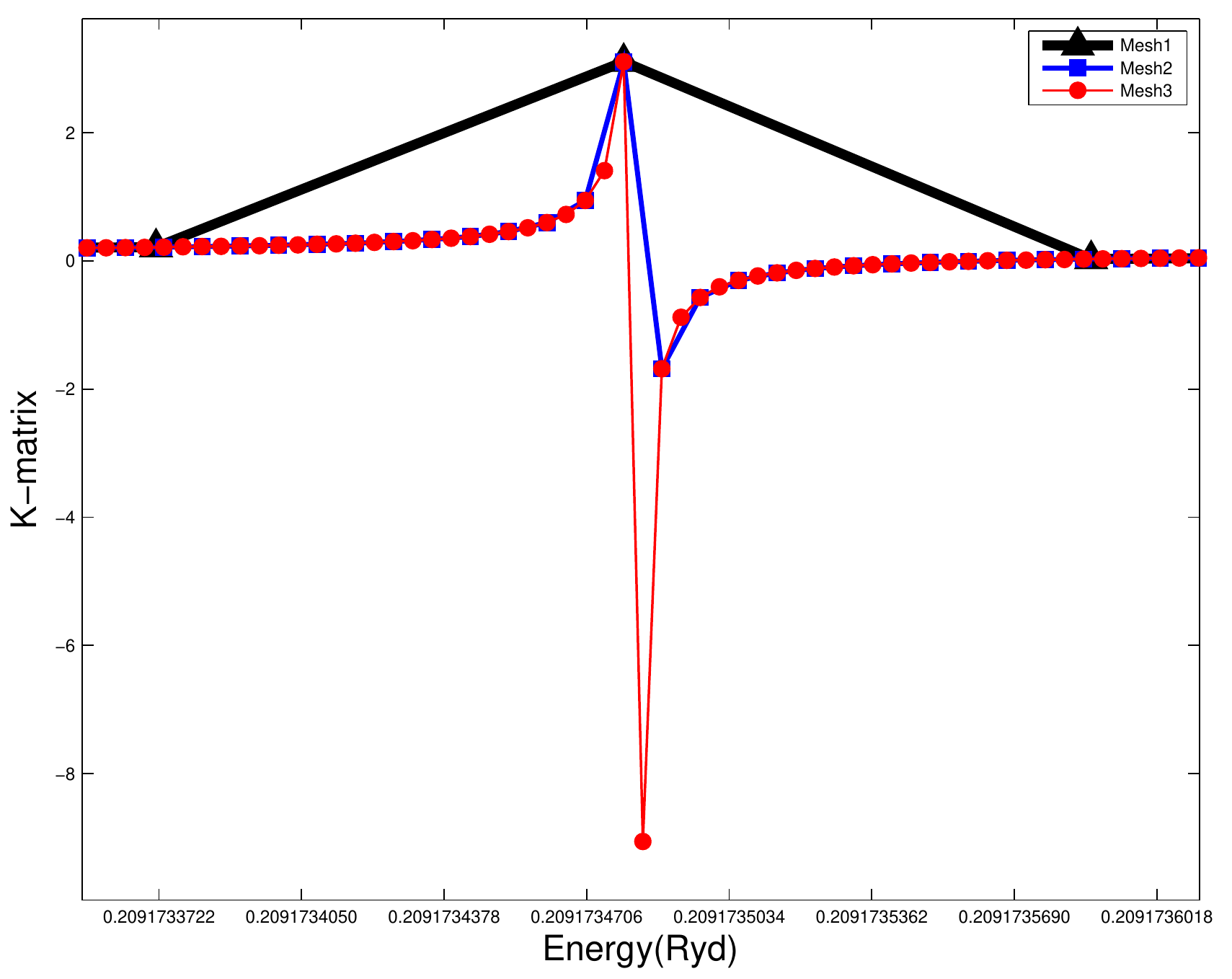} \caption{\Km-matrix as a function of energy for the first resonance in
Table~\ref{ResTable}, using three meshes, a coarse mesh, Mesh1, and two finer meshes, Mesh2 and Mesh3. The \Km-matrix sign change occurs only with Mesh2 and Mesh3.} \label{Graph32}
\end{figure}

\begin{figure}[!h]
\centering{}
\includegraphics
[scale=0.65] {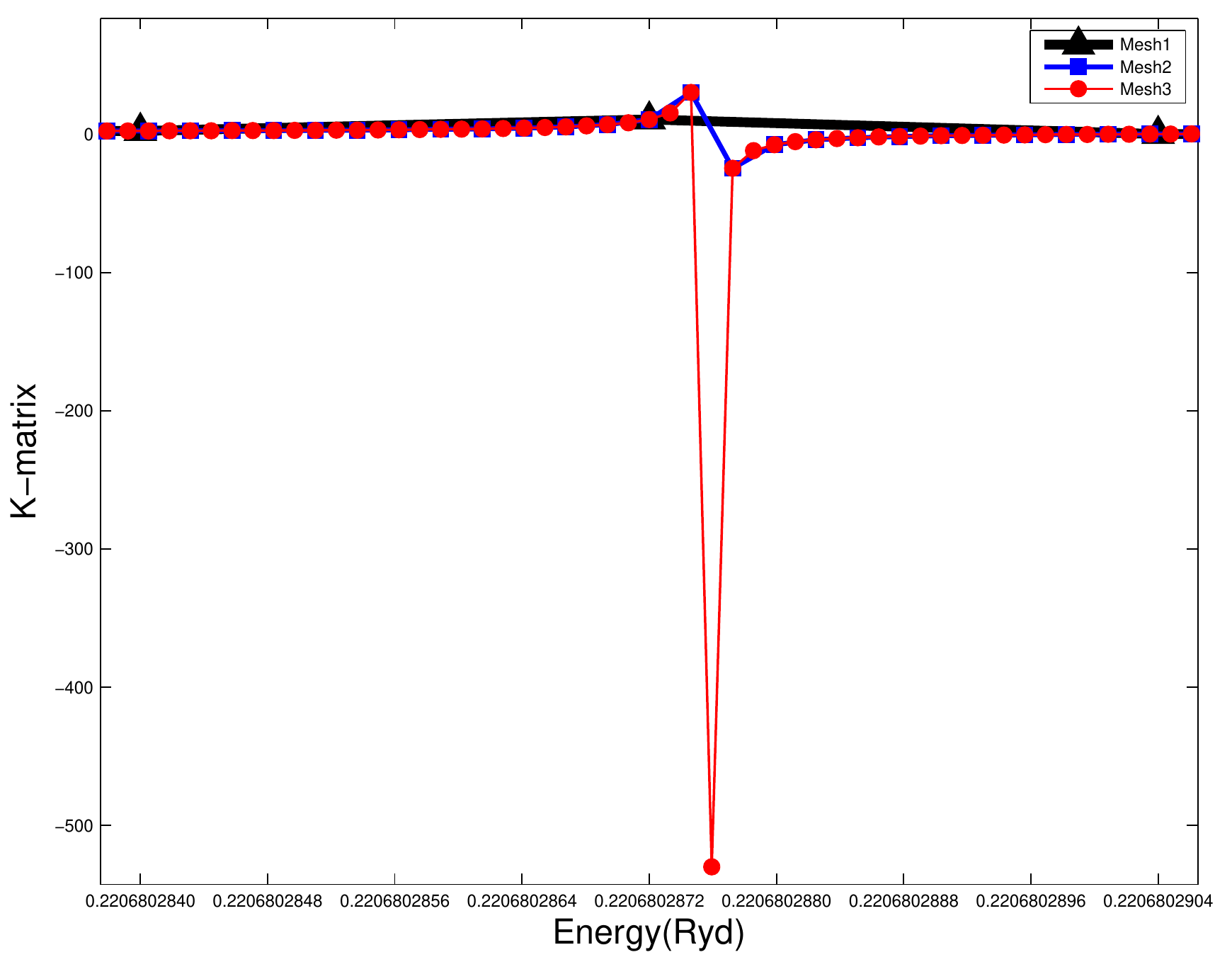} \caption{\Km-matrix as a function of energy for the second resonance in
Table~\ref{ResTable} using three meshes, a coarse mesh, Mesh1, and two finer meshes, Mesh2 and Mesh3. The \Km-matrix sign change occurs only with Mesh2 and Mesh3.} \label{Graph41}
\end{figure}

The process is demonstrated in Figures \ref{Graph32} and \ref{Graph41} which are based on two
examples from the C\II\ resonances \cite{SochiSCIIList2013} (refer to Table~\ref{ResTable} for
details). As seen in these figures, the local change in the magnitude of the gradient in the coarse
mesh (Mesh1) indicated the possible presence of a pole. The two figures also show that the two
finer meshes, Mesh2 with about 10 times more points, and Mesh3 with about 20 times more points,
have succeeded in finding the pole through the detection of sign change in the \Km-matrix although
the \Km-matrix profile is better delineated by Mesh3. For very narrow resonances care must be taken
to not compute the \Km-matrix too close to the pole as numerical noise can cause instability in the
derived resonance parameters.

A comparison between the \Km-matrix and \QB\ on the C\II\ resonances \cite{SochiThesis2012}
demonstrated that these methods produce virtually identical results. However, the \Km-matrix is
computationally superior in terms of the required computational resources, mainly CPU time.
Moreover, it is more powerful in detecting very narrow resonances which \QB\ cannot find.
Nevertheless, the \QB\ method is more general as it deals with multi-channel resonances, as well as
single-channel resonances, while the \Km-matrix method in its current formulation is restricted to
single-channel resonances.

\section{Conclusions} \label{Conclusions}

The resonance phenomenon plays very important role in the atomic scattering processes and
subsequent transitions. Several methods based on different theoretical backgrounds have been
proposed and used to find resonances and identify their parameters. In this article, we outlined
three methods for finding and analyzing resonances, including the recently developed \Km-matrix
method, which is highly efficient for investigating single-channel resonances near the ionization
threshold.

\newpage
\phantomsection \addcontentsline{toc}{section}{References} %
\bibliographystyle{unsrt}

\clearpage
\section[Derivation of \Km-Matrix Method]
{Appendix A: Using Lifetime Matrix to Investigate Single-Channel Resonances} \label{AppKmatrix}

In this Appendix we present the \Km-matrix method which is based on using the lifetime matrix $\Mm$
expressed in terms of the reactance matrix $\Km$ to investigate single-channel resonances.

In the case of single-channel states, \Mm, \Sm\ and \Km\ are one-element matrices. To indicate this
fact we annotate them with $\M$, $\Ss$ and $\K$. From Equations \ref{TDmatrixS2} and
\ref{SmatrixEq2}, the following relation can be derived
\begin{equation}\label{QmatrixK}
    \M = \frac{2}{1 + \K^{2}}  \D\K\E
\end{equation}
It is noteworthy that since $\K$ is real, $\M$ is real as it should be.

Smith \cite{Smith1960} has demonstrated that the expectation value of $\M$ is the lifetime of the
state, $\LTS$. Now if we consider a \Km-matrix with a pole superimposed on a background $\BGCK$
\begin{equation}\label{KwithBG}
    \K = \BGCK + \frac{\PKF}{\E - \Eo}
\end{equation}
then from Equation \ref{QmatrixK} we find
\begin{eqnarray}\label{QwithBG}
    \M(\E) &=& \frac{-2 \PKF}{(1 + \K^{2}) (\E - \Eo)^{2}} \nonumber \\
           &=& \frac {-2 \PKF} {(1 + \BGCK^{2}) (\E - \Eo)^{2} + 2 \BGCK \PKF (\E - \Eo) + \PKF^{2}}
\end{eqnarray}

The maximum value of $\M(\E)$ occurs when the denominator has a minimum, that is when
\begin{equation}\label{Emax}
    \E = \Eo - \frac {\BGCK \PKF} {1 + \BGCK^{2}}
\end{equation}
and hence
\begin{equation}\label{Qmax}
    \Mmax = - \frac {2 (1 + \BGCK^{2})} {\PKF}
\end{equation}
This reveals that by including a non-vanishing background the peak of $\M$ is shifted relative to
the pole position, $\E = \Eo$, and the peak value is modified. If we now calculate the full width
at half maximum, $\FWHMe$, by locating the energies where $\M = \frac{1}{2} \Mmax$ from solving the
quadratic
\begin{equation}\label{quadratic}
    (1 + \BGCK^{2}) (\E - \Eo)^{2}
    + 2 \BGCK \PKF (\E - \Eo)
    - \frac{\PKF^{2} (1 - \BGCK^{2})}{(1 + \BGCK^{2})}
    = 0
\end{equation}
we find
\begin{equation}\label{FWHMwithBG}
    \FWHMe = \frac {|2 \PKF|} {1 + \BGCK^{2}}
\end{equation}

The two main parameters in the investigation of autoionizing states are the resonance energy
position $\Er$ and its width $\RW$. For an energy point $\Ei$ with a \Km-matrix value $\SVKi$,
Equation~\ref{KwithBG} contains three unknowns, $\BGCK$, $\PKF$ and $\Eo$ which are required to
obtain $\Er$ and $\RW$, and hence three energy points at the close proximity of $\Eo$ are required
to identify the unknowns. Since the \Km-matrix changes sign at the pole, the neighborhood of $\Eo$
is detected by inspecting the \Km-matrix at each point on the energy mesh for sign reversal and
hence the three points are obtained accordingly. Now, if we take the three consecutive values of
$\K$
\begin{equation}\label{K123}
    \K_{i} = \BGCK + \frac{\PKF}{\E_{i} - \Eo} \verb|           | (i=1,2,3)
\end{equation}
and define
\begin{equation}\label{DELEKjk}
    \DELE jk = \E_{j} - \E_{k}
    \verb|       |  \&  \verb|       |
    \DELK jk = \K_{j} - \K_{k},
\end{equation}
then with some algebraic manipulation we find for $\Eo$, $\PKF$ and $\BGCK$,

\begin{equation}\label{E0}
\boxed{
    \Eo = \frac{ \E_{1} \DELK 12 \DELE 32 - \E_{3} \DELK 23 \DELE 21}
             {\DELK 12 \DELE 32 - \DELK 23 \DELE 21}
}
\end{equation}

\begin{equation}\label{PKF}
\boxed{
    \PKF
    = \frac{\DELK 12 \DELE 10 \DELE 20}{\DELE 21}
}
\end{equation}
and
\begin{equation}\label{BGCK}
\boxed{
    \BGCK = \K_{1} - \frac{\PKF}{\DELE 10}
}
\end{equation}
Finally, $\Er$ and $\RW$ can be computed from Equation~\ref{Emax} and Equation~\ref{FWHMwithBG}
respectively. Full details are given in \cite{SochiThesis2012}.

\end{document}